\documentstyle[prl,aps,twocolumn,floats]{revtex}
\begin{document}
\input epsf
\draft
\title{Turbulence and Multiscaling in the Randomly Forced Navier
Stokes Equation} 
\author{Anirban Sain$^1$, Manu$^2$ and Rahul
Pandit$^1$\cite{byjnc}} 
\address{$^1$Department of Physics, Indian Institute of Science,\\
Bangalore - 560 012, India and $^2$School of Physical Sciences,
Jawaharlal Nehru University, New Delhi - 110 067.}

\date{$\today$}

\maketitle

\begin{abstract} We present an extensive pseudospectral study
of the randomly forced Navier-Stokes equation (RFNSE) stirred
by a stochastic force with zero mean and a variance $\sim
k^{4-d-y}\/$, where $k\/$ is the wavevector and the dimension $d
= 3\/$. We present the first evidence for multiscaling of
velocity structure functions in this model for $y \geq 4\/$. We
extract the multiscaling exponent ratios $\zeta_p/\zeta_2\/$
by using extended self similarity (ESS), examine their
dependence on $y\/$, and show that, if $y = 4\/$, they are in
agreement with those obtained for the deterministically forced
Navier-Stokes equation ($3d$NSE). We also show that well-defined
vortex filaments, which appear clearly in studies of the
$3d$NSE, are absent in the RFNSE.
\end{abstract}



Kolmogorov's classic work (K41) on homogeneous, isotropic fluid
turbulence focussed on the scaling behavior of velocity
${\bf{v}}\/$ structure functions ${\cal {S}}_{p}(r)=\langle|{\bf
v}_{i}({\bf x} + {\bf r}) - {\bf v}_{i}({\bf x})|^p \rangle\/$,
where the angular brackets denote an average over the
statistical steady state \cite{k41}. He suggested that, for
separations $r \equiv |{\bf{r}}|\/$ in the {\em inertial range},
which is substantial at large Reynolds numbers $Re\/$ and 
lies between the forcing scale $L\/$ and the dissipation scale
$\eta_d\/$, these structure functions scale as ${\cal S}_{p} \sim
r^{\zeta_p}\/$, with $\zeta_p = p/3\/$. Subsequent experiments 
\cite{sreant+pram} have suggested instead that multiscaling obtains
with $p/3 > \zeta_p\/$, which turns out to be a nonlinear,
monotonically increasing function of $p\/$; this has also been
borne out by numerical studies of the three-dimensional
Navier-Stokes equation forced deterministically ($3d$NSE) at large 
spatial scales \cite{sreant+pram,chencao}. The determination 
of the exponents $\zeta_p\/$ has been one of the central, but 
elusive, goals of the theory of turbulence. One of the promising 
starting points for such a theory is the randomly forced 
Navier-Stokes equation (RFNSE) \cite{domart-FNS,YO1,YO2}, 
driven by a Gaussian random force whose spatial Fourier transform
${\bf{f}}({\bf{k}},t)\/$ has zero mean and a covariance
$<{\bf{f}}_i({\bf{k}},t) {\bf{f}}_j({\bf{k}}',t')> =
A k^{4-d-y} P_{ij}({\bf k}) \delta({\bf k+k'}) \delta(t-t')\/$;
here ${\bf{k}}, {\bf{k}}'\/$ are wave numbers, $t,t'\/$ times,
$i,j\/$ Cartesian components in $d\/$ dimensions, and
$P_{ij}({\bf k})\/$ the transverse projector which enforces the
incompressiblity condition. One-loop renormalization-group (RG)
studies of this RFNSE yield \cite{domart-FNS,YO1} a K41 energy
spectrum, namely, $E(k) \sim k^2 {S_2}(k) \equiv k^2 \langle |{\bf{v}}
({\bf{k}})|^2 \rangle \sim k^{-5/3}\/$, if we set $d = 3\/$ 
and $y = 4\/$; this has also been verified numerically 
\cite{YO2}. Nevertheless, these RG
studies have been criticised for a variety of reasons
\cite{weichman,eyink} such as using a large value for $y\/$
in a small-$y$ expansion and neglecting an infinity of marginal
operators (if $y = 4\/$). These criticisms of the {\em
approximations} used in these studies might well be justified;
but they clearly cannot be used to argue that the RFNSE is
{\em in itself}  inappropriate for a theory of turbulence. It is 
our purpose here to check if, indeed, the RFNSE is a good
starting point for such a theory. Specifically we want to test
whether structure functions in the RFNSE display the same
multiscaling as in the $3d$NSE for some value of $y\/$; if
they do, then we can argue that both equations are in the same
universality class and the RFNSE can, defensibly, be used to
develop a statistical theory of inertial-range multiscaling in
homogeneous, isotropic fluid turbulence.

To achieve this end we have carried out an extensive
pseudospectral study of the RFNSE and compared our results with
earlier numerical studies \cite{chencao,shprl} of
the $3d$NSE and experiments \cite{sreant+pram}. We find several
interesting and new results: We show that structure functions in
the RFNSE display multiscaling for $y \geq 4\/$. As in the $3d$NSE,
we obtain good estimates for ratios of the multiscaling
exponents, such as $\zeta_p/\zeta_2\/$, by using the
extended-self-similarity (ESS) procedure (Fig. 1)
\cite{shprl,benzi}; we obtain $\zeta_2\/$ from
$S_2(k)\/$ (Fig. 2).  Next we investigate the
$y$-dependence of $\zeta_p\/$ and find that it is close to the
$3d$NSE result (Fig. 3) for $y = 4\/$ at least for $p \leq 7\/$.
Thus the RFNSE should be a good starting point
for a theory of inertial-range multiscaling in the $3d$NSE
barring weak correction which do not affect ratios
like $\zeta_p/\zeta_2\/$ (see below). Furthermore we show that
the qualitative behaviors of the probability distributions
$P(\delta v_{\alpha}(r))\/$, where $\delta v_{\alpha}(r) \equiv
v_{\alpha}({\bf{x}}) - v_{\alpha}({\bf{x}} + {\bf{r}})\/$, are
similar in the two models (Fig. 4).  However, the shapes of
constant-$|\omega|\/$ surfaces, where $\omega\/$ is the vorticity, 
are markedly different (Fig. 5); the stochastic force destroys 
well-defined filamentary structures that obtain in $3d$NSE studies. 
This has implications for the She-Leveque \cite{sl} formula 
for $\zeta_p\/$ as we discus below.

We use a pseudospectral method to solve the RFNSE numerically on
a $64^3\/$ grid with a cubic box of linear size $L = 2\pi\/$ and
periodic boundary conditions; we have checked in representative
cases that our results are unchanged if we use an $80^3\/$ grid
or aliasing. Aside from the stochastic forcing, our numerical
scheme is the same as that used in  Ref.\cite{shprl}. Our
dissipation term, which is $(\nu + \nu_H k^2)k^2
{\bf{v}}({\bf{k}})\/$ in wave-vector ($k\/$) space includes both
the viscosity $\nu\/$ and the hyperviscosity $\nu_H\/$; the
exponents $\zeta_p\/$ are unaffected by $\nu_H\/$ if $\nu > 0\/$
\cite{shprl,chencaoshe}. Our numerical study of the RFNSE differs
from conventional studies of the $3d$NSE in two important ways:
(1) For a fixed grid size we can attain higher Taylor-microscale
Reynolds numbers $Re_{\lambda}\/$ in the RFNSE, and hence a
larger inertial range, than in the $3d$NSE ($Re_{\lambda} \simeq
120\/$ compared to $Re_{\lambda} \simeq 22\/$ in our study), as
noted earlier \cite{YO2} for $y = 4\/$.  (2) This advantage is
reduced somewhat by the need to average statistical observables
longer in the RFNSE than in the $3d$NSE. In the latter case it
normally suffices to average over a few box-size eddy turnover
times $\tau_{L}\/$; this is not enough for the RFNSE since (a)
$Re_{\lambda}\/$ fluctuates strongly over time scales
considerably larger than $\tau_{L}\/$ (inset of Fig. 2)
and (b) the length of the ${\bf f(k},t)\/$ time series required
to obtain a specified variance for the stochastic force turns
out to be quite large (the time series must be of length $\simeq
6 \tau_{L}\/$ to achieve the given variance within $1-2\%\/$).
Thus, in our studies, we have collected data for averages over
$25-33\tau_{L}\/$ (for different values of $y\/$ ), after initial 
transients have been allowed to decay  
(over times $\simeq 10-20\tau_{L}\/$). 
Our $\tau_{L} \simeq 10 \tau_I \/$, the integral-scale
time used in some studies \cite{menvin}; $\tau_I \equiv
L_I/v_{rms}\/$, where the integral scale $L_I \equiv [\int dk k
E(k) /\int dk E(k)]^{-1}\/$ and $v_{rms}\/$ is the 
root-mean-square velocity.

\begin{figure}[t]
\epsfxsize=18pc
\epsfysize=16pc
\centerline{\epsfbox{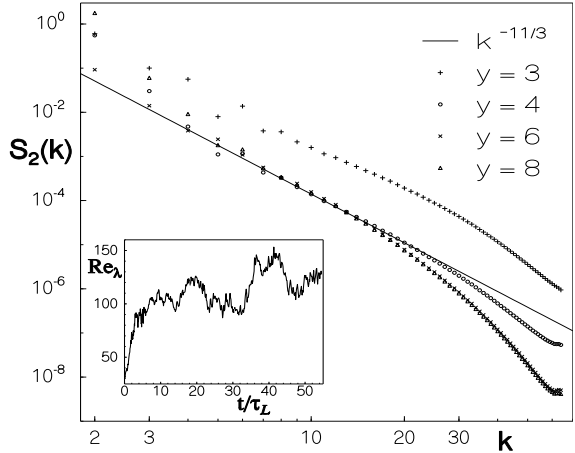}}
\caption{Log-log plots (base 10) of $S_2(k)\/$ versus $k\/$ for
different values of $y\/$. The line indicates the K41 result 
$S_2(k)\sim k^{-11/3}\/$. The inset shows a repesentative plot of 
$Re_{\lambda}\/$ versus time($t\/$) for $y=4\/$. }
\label{fig1}
\end{figure}

We begin by investigating the inertial-range scaling of the 
$k$-space structure function $S_2(k) \sim k^{-\zeta'_2}\/$. Given this 
power-law form, the exponent $\zeta'_2\/$ is easily seen to be
related to the $r-$space inertial-range exponent $\zeta_2\/$ by
the formula $\zeta_2 = \zeta'_2 - 3\/$. Our data in Fig. 1 
 for $4 \leq y\/$ are consistent with $\zeta'_2 = 11/3\/$ 
(i.e., the K41 value since $E(k) \sim k^2 {S_2}(k) \sim k^{-5/3}
\/$). For $y = 4\/$ this 
result has been reported earlier \cite{YO2}. The $y-$independence 
of $\zeta'_2\/$ above some critical $y_c\/$ (our data suggest $y_c
\simeq 4\/$) is theoretically satisfying since the variance of
the stochastic force in the RFNSE rises rapidly at small $k\/$,
so we might expect that, for sufficiently large $y\/$, it
approximates the conventional deterministic forcing of the
$3d$NSE at large spatial scales.  This point of view has been
explored in the $N \rightarrow \infty\/$ limit of an
$N-$component generalization of the RFNSE \cite{weichman};
however, this study suggests $\zeta'_2 = 7/2\/$ for $y_c = 4
\leq y\/$; given our error bars (Table 1) it is difficult to
distinguish this from the $O(y)\/$ RG prediction $\zeta_2' =
11/3\/$ though our data are closer to the latter. 
For $0 < y \leq 3\/$ both the one-loop RG \cite{YO1}
and the $N \rightarrow \infty\/$ theory \cite{weichman} predict
$\zeta'_2 \sim 1 + 2y/3 + O(y^2)\/$, which is in fair agreement
with our numerical results, especially for small $y\/$ (Table 1).
We note in passing that, for $0 < y < 4\/$, there is
no {\em invariant} energy cascade as in conventional K41: The
dominance of dissipation at large $k\/$ does lead to an energy
cascade, but the energy flux depends on the length scale
$r\/$; specifically $\Pi(r) \approx A r^{y-4}\/$, with
$A\/$ the scale-independent part of the variance of the
stochastic force in the RFNSE. A K41-type argument \cite{frish}
now yields an energy-transfer rate $\sim {<\delta v_{r}^{3}>}/r
\sim r^{(y-4)}\/$, whence ${\cal S}_3(r) \sim r^{(y-3)}\/$ and,
if we assume simple scaling as in K41, ${\cal S}_2(r) \sim
r^{(y-3)2/3}\/$, i.e., $\zeta_2' = 1 + 2y/3\/$, which is same as
the $O(y)\/$ RG prediction mentioned above. This formula breaks
down for $y < 0\/$; however, the RG predicts correctly that the
linear-hydrodynamics result obtains in this regime. 

Several precautions must be taken to ensure that systematic
errors do not affect the determination of $\zeta_2'\/$.
If $k_{max}\/$ is the largest wave-vector magnitude in our
numerical scheme, we find that $L_I k_{max}\/$ decreases with
decreasing $y\/$; this shortens the inertial range of the energy
spectrum which can be used to obtain $\zeta_2'\/$.
The lower the value of $y\/$ the more difficult it is to obtain
a dissipation range free of systematic, finite-resolution
errors.  For $y < 4\/$, we define $k_d \equiv \eta_d^{-1}\/$ to
be the inverse length scale at which the energy-transfer time
$t_{r} \sim (r/v_{r}) \sim [A r^{(y-6)}]^{1/3}\/$
equals the diffusion time $t_D \sim [\nu k^2 + \nu_H
k^4]^{-1}\/$; this yields
\begin{equation}
\nu_0 k_d^2 + \nu_h k_d^4 = [A k_d^{6-y}]^{1/3},
\end{equation}
which when solved numerically shows that, for fixed $A\/$,
$k_d\/$ increases as $y\/$ decreases (Table 1).  It is important to
recognize that statistical steady states, with ill-resolved
dissipation ranges that do not have a decaying
tail \cite{shprl}, can be obtained by adjusting the amplitude
$A\/$. In such cases $ k_d \gg k_{max}\/$ and we get spurious
results for $\zeta_2'\/$. We find that, if we increase the
hyperviscosity $\nu_H\/$, $k_d\/$ is sufficiently close to
$k_{max}\/$ that we can resolve both inertial and dissipation
ranges and obtain reliable values for $\zeta_2'\/$. Table 1
shows the range over which we fit our data for $S_2(k)\/$. 
Since our data for $\zeta_2'\/$ indicate that $y_c\simeq 4\/$, 
we investigate multiscaling only for $y \geq 4\/$.

\begin{figure}[htb]
\epsfxsize=17pc
\epsfysize=12pc
\centerline{\epsfbox{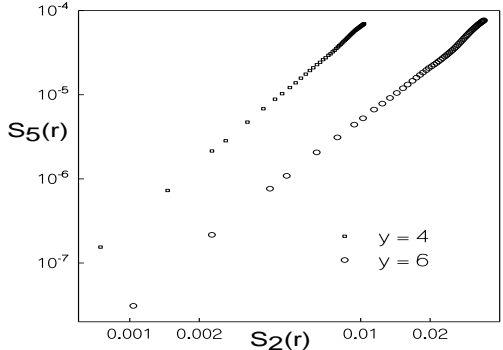}}
\caption{Log-log plots (base 10) of ${\cal{S}}_5(r)\/$ versus
${\cal{S}}_2(r)\/$ illustrating the ESS procedure.} 
\label{fig2}
\end{figure}
\begin{figure}[htb]
\epsfxsize=17pc
\epsfysize=13pc
\centerline{\epsfbox{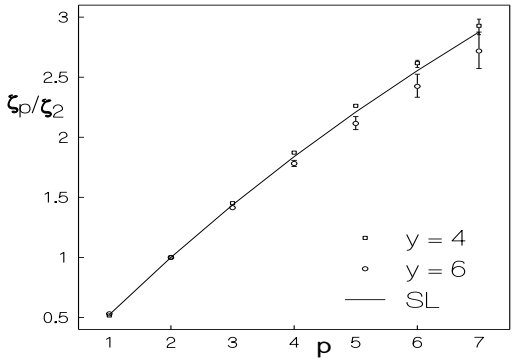}}
\caption{Inertial-range exponent ratios $\zeta_p/\zeta_2\/$ 
versus $p\/$ for the RFNSE with $y=4\/$ and $6\/$ (extracted
from plots like Fig.2). The line indicates the SL formula.}
\label{fig3}
\end{figure}

Our data for $\zeta_2'\/$ in Table 1 suggest that naive
estimates for the multiscaling exponents $\zeta_p\/$ require
longer inertial ranges than are available in our studies. However, 
we find that, as in the $3d$NSE, the
extended-self-similarity (ESS) procedure \cite{chencao,shprl,benzi} 
can be used fruitfully here to extract the exponent ratios 
$\zeta_p/\zeta_q\/$ from the
slopes of log-log plots of ${\cal{S}}_p(r)\/$ versus
${\cal{S}}_q(r)\/$ (see Fig. 2 for $p = 5\/$ and $q = 2\/$)
since this extends the apparent inertial range.  The ratios
$\zeta_p/{\zeta_2}\/$ that we obtain from such ESS plots are
compared in Fig. 3 with the She-Leveque (SL) formula
\cite{sl}, which provides a convenient parametrization
for the experimental values for $\zeta_p\/$. (If we
assume the power-law form for $S_2(k)\/$ in the
inertial range (Fig. 1), we get $\zeta_2 = \zeta'_2
- 3\/$ and thence all the exponents $\zeta_p\/$.) Figure 3 shows
clearly that, with $y = 4\/$, our RFNSE exponent ratios lie very 
close to those for the $3d$NSE and, to this extent, these two
models are in the same universality class. The exponent ratios
for $ y < 4\/$ lie away from the $3d$NSE values as we might have
anticipated from our results for $S_2(k)\/$ (Fig. 1).
These results also lead to the expectation that, if $y\/$ is
sufficiently large, the exponent ratio $\zeta_p/\zeta_2\/$ 
should become
independent of $y\/$, for we find that $\zeta_2(y=4) \simeq
\zeta_2(y = 6)\/$. However, for $p > 3\/$, our data for
$\zeta_p/\zeta_2(y=6)\/$ fall systematically below those for
$\zeta_p/\zeta_2(y=4)\/$ or the SL line. We also find that that the
probabilty distributions of $P(\delta v_r)\/$ (Fig. 4) have 
non-Gaussian tails for $r\/$ in the dissipation range; and for
$y > 4\/$ the deviations from a Gaussian distribution increase
systematically with $y\/$. Thus, at least at the resolution of our 
calculation, it seems that the RFNSEs with $y = 4\/$ and $y =
6\/$ are in different universality classes. However, we wish to
point out that our data for $y = 6\/$ are more noisy than those
for $y = 4\/$, so longer runs with finer grids might well be
required to settle this issue conclusively.

\begin{figure}
\epsfxsize=17pc
\epsfysize=13pc
\centerline{\epsfbox{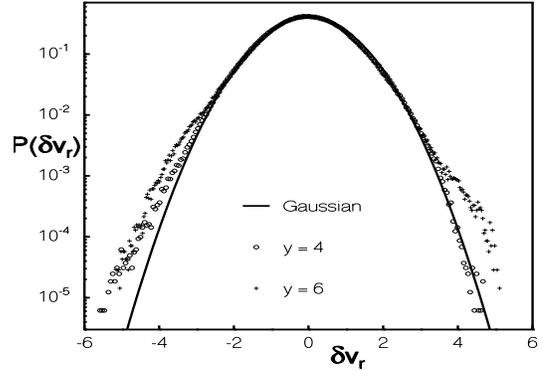}}
\caption{Semilog plots of the distribution $P(\delta v_r)\/$
(i.e., $P(\delta v_{\alpha}(r))\/$ averaged over $\alpha\/$) for 
$r\/$ in the dissipation range and $y=4\/$ and $6\/$. A Gaussian 
distribution is shown for comparison.}
\label{fig4}
\end{figure}

Strictly speaking the RFNSE with $y=4\/$ falls in the same 
universality class as the $3d$NSE only in the ESS sense. For 
arbitrary $y\/$ the energy flux through the $k^{th}\/$ shell is
${\Pi_k}\equiv \Pi(r=k^{-1}) \sim \int_{1/L}^{k}
\langle |{\bf{f}}({\bf{k}})|^{2} \rangle {d^3}k\/$, 
where $r\/$ is in the inertial-range and we have used Novikov's 
theorem \cite{frish}, i.e., $\langle {\bf{f}}({\bf{k}}) 
\cdot {\bf{v}}({\bf{-k}})\rangle\/ \sim \langle |{\bf{f}}({\bf{k}})
|^{2}\rangle\/$.
For $y > 4\/$, $\Pi_k\/$ saturates to a constant for $kL\gg 1\/$; 
but for $y = 4\/$, $\Pi_k \sim \; \log (kL)\/$ in the RFNSE. This 
is to be contrasted, with the $3d$NSE where $\Pi_{k} = constant\/$. 
Thus the inertial-range behaviors of all correlation functions in 
the two models are not the same.
%
%
A K41-type dimensional analysis suggests that for
$y=4\/$ the energy flux ${\Pi_k} \sim <\delta v_{r}^{3}>/r 
\sim \log(r/L)\/$; if we further assume that there is no 
multiscaling, then ${\cal S}_{p}(r) \sim [r \log(r/L)]^{p/3}\/$. 
Multiscaling will clearly modify this simple prediction; 
but some weak deviation from the Von-Karman-Howarth form 
${\cal S}_{3}(r) \sim r\/$ must remain, since the standard 
derivation of the Von-Karman-Howarth 
relation \cite{frish} does not 
go through \cite{longpap} with the RFNSE result for $\Pi_k\/$. 
To the extent that our data show that the ESS procedure works
for the RFNSE, it seems that these weak deviations cancel when 
we consider the ratios of structure functions; and, as noted 
above, for $y=4\/$ the exponent ratios $\zeta_{p}/\zeta_{2}\/$ 
agree with the SL result for the $3d$NSE.


\widetext
\begin{table*}[t]
\caption{The dissipation-scale wavenumber $k_d\/$ (determined
from Eq. 1), the integral-scale wavenumber $k_I \equiv
L_I^{-1}\/$, the apparent inertial range over which we fit our
data for $S_2(k)\/$, the hyperviscosities $\nu_H\/$, the
exponent $\zeta_2'\/$ that we compute, and its $O(y)\/$ RG
value, for $1 \leq y \leq 4\/$. The viscosity $\nu\/$ is $5 \times
10^{-4}\/$ in all these runs which use a $64^3\/$ grid.}
\label{table1}
\begin{tabular}{|c|c|c|c|c|c|c|}
\hline
$y$ & $k_d$ & $k_I$ & Fitting Range & $\nu_{H}$ & $\zeta_2'$
this  study & $\zeta_2'$ from $O(y)\/$ RG  \\ 
\hline
4 & 49 & 1.16 & $(0.1 - 0.5) k_d$ & $10^{-6}$ & $3.6 \pm 0.1$
& $\simeq 3.67\/$ \\
3 & 38.7 & 1.90 & $(0.16 - 0.52) k_d$ & $3 \times 10^{-6}$ & $3.0 \pm 0.1$
& $\simeq 3\;\;\;\;\/$ \\
2 & 35.0 & 5.90 & $(0.17 - 0.63) k_d$ & $8 \times 10^{-6}$ &
$2.3 \pm 0.1$ & $\simeq 2.33$ \\
1 & 35.4 & 10.3 & $(0.2 - 0.7) k_d$ & $8 \times 10^{-6}$ & $1.6
\pm 0.15\/$ & $\simeq 1.67$ \\
\end{tabular}
\end{table*}
\narrowtext

\begin{figure}[h*]
\epsfxsize=16pc
\centerline{\epsfxsize=9pc \epsfbox{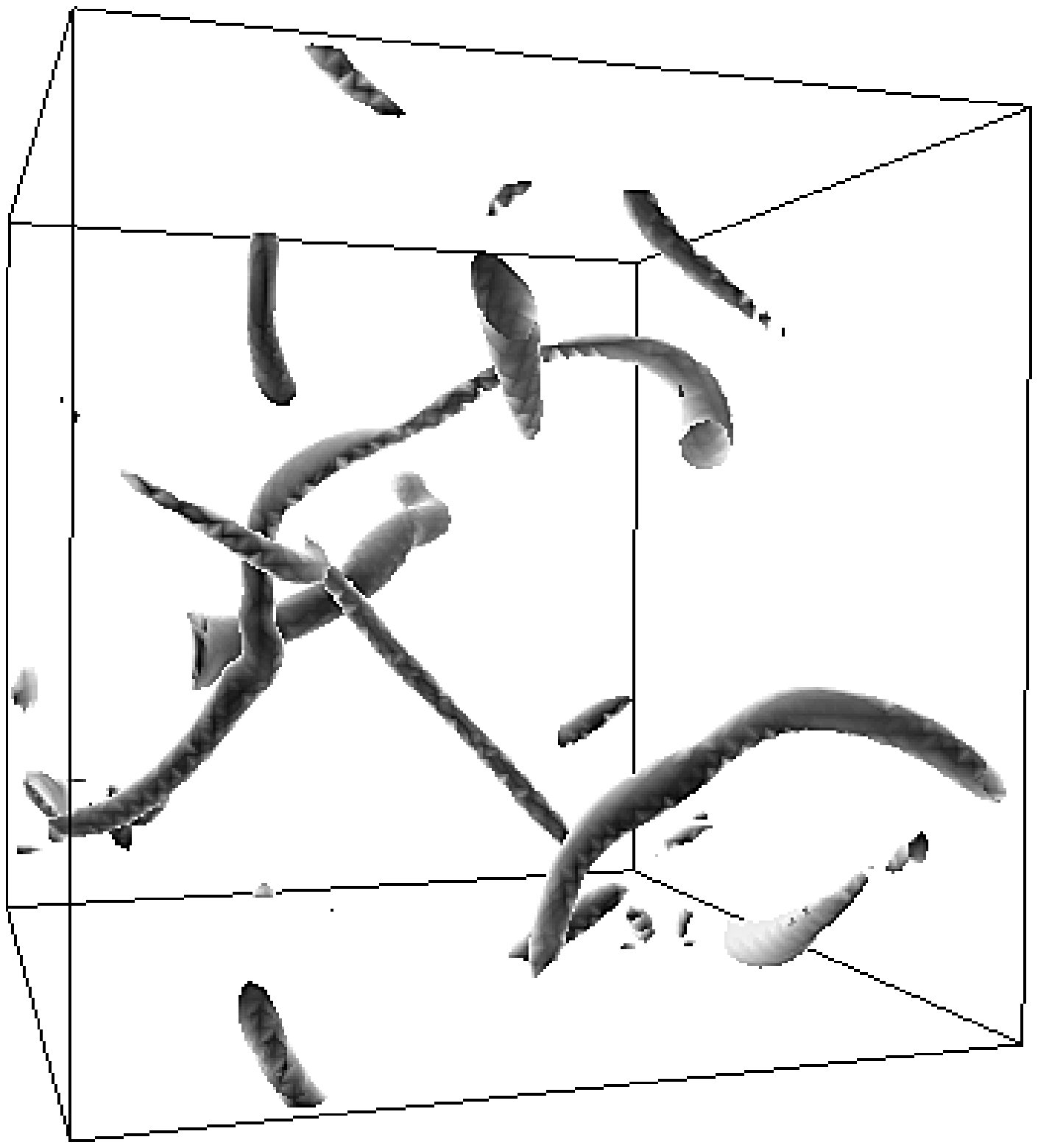}  
\hspace{.3cm}
\epsfxsize=7.5pc \epsfbox{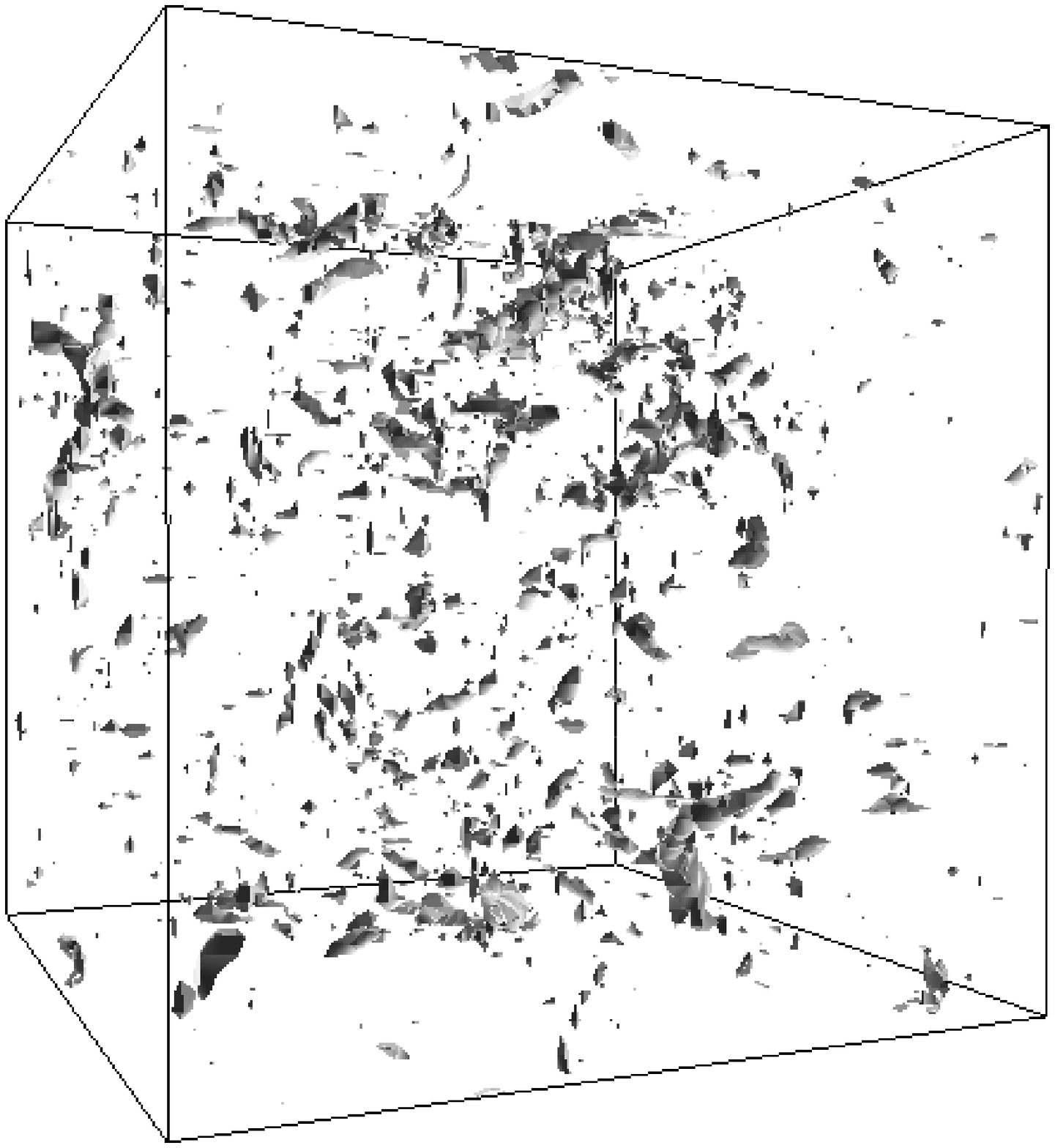}}
\vspace{-.25cm}
\caption{Iso-$|\omega|\/$ surfaces obtained from instantaneous
snapshots of the vorticity fields showing filaments for the
$3d\/$NSE (left) and no filaments for the RFNSE with 
$y = 4\/$ (right).}
\label{fig5}
\end{figure}


Filamentary structures (Fig. 5) \cite{siggia-orz-ourev} in 
iso-$|\omega|\/$ plots have been used as important  
ingredients in phenomenological models for multiscaling in 
fluid turbulence. For example, the SL formula \cite{sl} is obtained
by postulating a hierarchical relation among the moments of the 
scale-dependent energy dissipation; this yields a difference equation for 
the exponents $\tau_p\/$, which are simply related to the exponents 
$\zeta_p\/$; one of the  crucial boundary conditions used to solve this equation 
requires the codimension of the most intense structures. If these are 
taken to be vorticity filaments, their codimension is $2\/$ and one gets the 
SL formula. Filaments have been observed in experiments also \cite{douady}.
We have shown above that the exponent ratios $\zeta_p/\zeta_2\/$ that we obtain from 
the RFNSE with $y = 4\/$ agree with the SL formula. One might expect, therefore,
that filamentary structures should appear in iso-$|\omega|\/$ plots for 
the RFNSE. However, this is not the case as can be seen from the representative
plot shown in Fig. 5. The stochastic forcing seems to destroy the well-defined 
filaments observed in the $3d$NSE {\em without changing the multiscaling
exponent ratios}. Therefore, the existence of vorticity filaments is not crucial for 
obtaining these exponents, which is perhaps why simple shell models 
\cite{shprl,pisa} also yield good estimates for $\zeta_p\/$.

In summary, then, we have shown that the RFNSE with $y=4\/$ exhibits
the same multiscaling behavior as the 3dNSE, at least in the ESS sense.
Probability distributions like $P(\delta v_r)\/$ (Fig. 4) are 
also qualitatively similar in the two models, in so far as they show 
deviations from Gaussian distributions for $r\/$ in the dissipation 
range. It would be interesting to see if the RFNSE model can be
obtained as an effective, inertial-range equation for fluid turbulence.
 We have tried to do this by a coarse-graining procedure that has been 
used \cite{ks} to map the Kuramoto-Sivashinsky(KS) equation onto the 
Kardar-Parisi-Zhang (KPZ) equation; however, it turns out that the 3dNSE 
$\rightarrow\/$ RFNSE mapping, if it exists, is far more subtle than the
KS $\rightarrow\/$ KPZ mapping as we discuss elsewhere \cite{longpap}. 


We thank S. Ramaswamy for discussions, Ashwin Pande for help with
Fig. 5, CSIR (India) for
support, and SERC (IISc, Bangalore) for computational resources.

\end{document}